\documentclass[prl,showpacs,footinbib,twocolumn,final,superscriptaddress]{revtex4}

\usepackage{amsfonts,amsmath,amssymb,ifpdf,epsfig,graphicx,color}

\newcommand{\Eq}[1]{Eq.~(\ref{#1})}

\newcommand{\boldfx}{x}

\newcommand{\Tph}{\tau_{\varphi}}
\newcommand{\glead}{g_{\rm lead}}
\newcommand{\transmission}{T_{\rm cont}}
\newcommand{\gcontact}{g_{\rm cont}}
\newcommand{\TphH}{\tau_{H}}
\newcommand{\TTh}{\tau_{\text{Th}}}
\newcommand{\Tdw}{\tau_{\textrm{dw}}}
\newcommand{\ETh}{E_{\text{Th}}}
\newcommand{\tauT}{{\tau_T}} 
\newcommand{\envelope}{\textrm{en}}
\newcommand{\Tmin}{T_{\text{dil}}}
\newcommand{\Tmax}{T_{\text{ph}}}

\begin{document}

\title{Dimensional Crossover of the Dephasing Time in Disordered
  Mesoscopic Rings}

\author{M. Treiber}

\author{O.M. Yevtushenko}

\author{F. Marquardt}

\author{J. von Delft}

\affiliation{Ludwig Maximilians University, Arnold Sommerfeld Center and
             Center for Nano-Science, Munich, D-80333, Germany}

\author{I.V. Lerner}

\affiliation{School of Physics and Astronomy, University of Birmingham,
             Birmingham, B15 2TT, UK}

\begin{abstract}
  We study dephasing by electron interactions in a small disordered
  quasi-one dimensional ($1D$) ring weakly coupled to leads.  We use
  an influence functional for quantum Nyquist noise to describe the
  crossover for the dephasing time $\Tph (T)$ from diffusive or
  ergodic $1D$ ($ \Tph^{-1} \propto T^{2/3}, T^{1}$) to $0D$ behavior
  ($\Tph^{-1} \propto T^{2}$) as $T$ drops below the Thouless energy.
  The crossover to $0D$, predicted earlier for $2D$ and $3D$ systems,
  has so far eluded experimental observation.  The ring geometry holds
  promise of meeting this longstanding challenge, since the crossover
  manifests itself not only in the smooth part of the
  magnetoconductivity but also in the amplitude of
  Altshuler-Aronov-Spivak oscillations.  
  This allows signatures of  dephasing in the ring to be cleanly 
  extracted by filtering out those of the leads.
\end{abstract}

\date{May 3, 2009}


\maketitle

Over the last twenty years numerous
theoretical~\cite{AAK,Fukuyama1982,AA-review,SIA,AleinerEtAl,SashaMontamb,FlorianJan-1,FlorianJan-2}
and experimental~\cite{Marcus:93,yacoby94,reulet95,DotExper,Esteve:2000a,NetExper}
works have
studied the mechanism of dephasing in electronic transport and its
dependence on temperature $T$ and dimensionality in disordered
condensed matter systems.  At low temperatures dephasing is mainly due
to electron interactions, with the dephasing time $\Tph (T)$ increasing as
$T^{-a}$ when $T \to 0$.

The dephasing time controls the scale of a negative weak
localization (WL) correction to the magnetoconductivity, and
(under certain conditions) the magnitude of universal conductance
fluctuations (UCFs).  If $T$ is so low that $\Tph$ exceeds $\TTh =
{\hbar} / \ETh$, the time required for an electron to cross
(diffusively or ballistically) a mesoscopic sample ($\ETh$ is the
Thouless energy), UCFs become $T$-independent. This leaves WL as the
only tool to measure the $T$-dependence of dephasing in mesoscopic
wires or quantum dots at very low $T$.
For quantum dots, a dimensional
crossover was predicted \cite{SIA} from $\Tph \propto T^{-1}$, typical
for a $2D$ electron gas \cite{AAK}, to $\Tph \propto T^{-2}$ when the
temperature is lowered into the $0D$ regime,
\begin{align}\label{1}
\hbar/   \Tph  \ll  T \ll \ETh \, ,
\end{align}
where the coherence length and the thermal length are both larger than
the system size, independent of geometry and real dimensionality of
the sample.  Although the $ \Tph \propto T^{-2} $ behavior is quite
generic, arising from the fermionic statistics of conduction
electrons, experimental efforts \cite{DotExper} to observe it
have
so far been unsuccessful.  The reasons for this are
unclear. Conceivably dephasing mechanisms other than electron
interactions were dominant, or the regime of validity of the $0D$
description had not been reached.  In any case, other ways of testing
the dimensional crossover for $\Tph $ are desirable.

Here we study dephasing in a quasi-$1D$
mesoscopic ring weakly coupled to two well-conducting leads
through narrow point contacts.
We find a dimensional crossover for $\Tph (T)$ from
diffusive or ergodic $1D$ ($\propto T^{-2/3},T^{-1}$) to $0D$ 
($\propto T^{-2}$) behavior as
$T$ is decreased below $\ETh$, and propose a detailed
experimental scenario for observing this behavior. It reveals itself
not only via the WL corrections to the smooth part of
magnetoconductivity, but also via the amplitude of the
Altshuler-Aronov-Spivak (AAS) oscillations \cite{MagnetoOsc} that
result from closed trajectories with a non-zero winding number
acquiring the Aharonov-Bohm phase.  
Under suitable conditions, discussed below,
the magnitude of AAS oscillations will be independent of
dephasing in the leads.
Thus, the ring geometry provides a more
promising setup for the observation of the dimensional crossover.
than $2D$ or $3D$ settings.

\textit{Dephasing in weak localization.} The WL correction
to the conductivity
is governed by coherent back-scattering of the electrons from
static disorder and, to the lowest order, is due to the
enhancement of the return probability caused by constructive
interference of two time reversed trajectories, described by the
so-called Cooperon ${\cal C} $ \cite{GLKh,MontBook}.  In this order,
the WL correction to the conductivity, in units of the Drude
conductivity $\sigma_0$, is
given by \cite{DK:84}:
\begin{equation}
\label{WL-sigma}
\Delta g = \frac{  \Delta \sigma}{\sigma_0}
 = - \frac{1}{\pi \nu } \int_0^{\infty}
{\rm d} t' \,  {\cal C}(t') \, %e^{ -{\cal F}( t'/\Tph ) } \,
\, .
\end{equation}
Here $ \nu $ is the electron density of states per spin at the Fermi
surface, and $ \hbar = 1 $ henceforward.  Dephasing limits
the scale of this contribution and effectively results in the
suppression of the Cooperon at long times:
\begin{equation}
  {\cal C}(t) \equiv {\cal C}_0(t) 
\exp \left[ - t / \TphH  - t / \Tdw  - {\cal F}(t) \right]  \, .
\end{equation}
We consider here low temperatures where the phonon contribution to
dephasing is negligible and three main sources contribute to the
  Cooperon decay with time: an applied magnetic field $H$,
  characterized by the time scale $\TphH$ \cite{AA-magnetores}; the
  leakage of particles from the ring, characterized by the dwell time
  $ \Tdw $ \cite{BeeMcCannLerner}; and electron interactions,
  whose effects can be described in terms of the decay function $
  {\cal F}(t) $ \cite{AAK,AA-review}, which grows with time and may be
  used to define a dephasing time via ${\cal F} (\Tph) = 1$.
  
  $ {\cal F}(t) $ can be obtained using the influence functional
  approach \cite{SashaMontamb,FlorianJan-1}, which gives % Montamb
  results for the magnetoconductivity that are practically equivalent
  to those originally obtained in \cite{AAK}. Roughly speaking, an
  electron traversing a random walk trajectory $\boldfx(t_1)$ of
  duration $t$ acquires a random phase $\varphi_t = \int_0^t
  \textrm{d} t_1 V( \boldfx(t_1),t_1) $ due to the random potential
  $V$ describing the Nyquist noise originating from electron
  interactions; the variance of this phase, averaged over all closed
  random walks (crw), gives the decay function, ${\cal F} (t) =
  \frac{1}{2} \langle \delta \varphi_t^2 \rangle_{\textrm{crw}}$.  A
  careful treatment \cite{SashaMontamb,FlorianJan-1} gives % Montamb
\begin{equation}
\label{DecFunc}
  {\cal F}(t) = \int_{0}^{t} \!\! {\rm d^2} t_{1,2}
    \Bigl\langle
         \overline{V V} (\boldfx_{12}, t_{12})
       - \overline{V V} (\boldfx_{12}, \bar t_{12})
    \Bigr\rangle_{\textrm{crw}},
\end{equation}
where $ \boldfx_{12} \equiv \boldfx(t_1) - \boldfx(t_2) $; $ t_{12}
\equiv t_1 - t_2 $; $ \bar t_{12} \equiv t_1 + t_2 - t $.  The noise
is assumed Gaussian, with correlation function
\begin{equation}\label{Q-corr}
  \overline{V V} (\boldfx ,t) = (2 e^2 \, T / \sigma_0 A) \, 
Q(\boldfx) \ \delta_T(t) \, .
\end{equation}
Here $A$ is the wire's cross-sectional
area, the diffuson $ Q(\boldfx) $ 
is the time-averaged solution of the diffusion equation and $
\delta_T(t) $ is a broadened $ \delta$-function of width $\tauT \simeq
1/T$ and height $\simeq T$, given by   \cite{FlorianJan-1,FlorianJan-2}
\begin{equation}\label{delta-T}
  \delta_T(t) =  \pi T w(\pi \, T \, t) \, , \quad
  w(y) = \frac{y \coth(y) - 1}{\sinh^2(y)} \, .
\end{equation}
This form takes into account the Pauli principle in a quantum
description of Nyquist noise and reproduces the results
\cite{AleinerEtAl} of leading order perturbation theory in the interaction
for $\Delta g$.  The broadening of $\delta_T (t)$ is the central
difference between quantum noise and the classical noise considered in
previous treatments \cite{AAK,SashaMontamb}, which used a sharp
$\delta(t)$ function instead. Note that \Eq{DecFunc} is free from IR
singularities, because the $\boldfx$-independent part of $\overline{V
  V}$ (the diffuson ``zero mode'') does not contribute to ${\cal F}$.

\begin{figure}[t]
\ifpdf
  \includegraphics[width=\columnwidth]{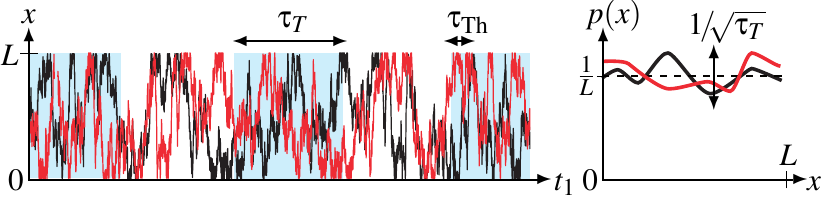}
\else
  \epsfig{file=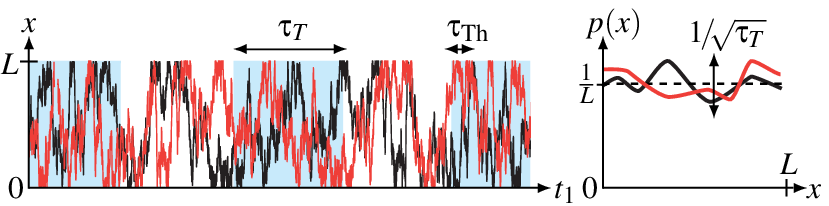,width=\columnwidth}
\fi
\caption{(Color online) Left: A pair of time-reversed diffusive trajectories
  exploring ergodically a region of size $L$. The fluctuating noise
  potential is frozen during time-intervals (indicated by shading)
of duration $\tauT=1/T$, sketched here to be $\gg \TTh$.
Right: The density $p(x)$ of points $x$ visited
by a particular trajectory during the time-interval $\tauT$ %\gg\TTh$
fluctuates around $1/L$, with fluctuations $\delta p \sim 1 / \sqrt{\tauT}$.}
\label{Figure1}
\end{figure}

\emph{Qualitative picture.}  We begin with a qualitative
  discussion of dephasing in an isolated quasi-$1D$ system of size
  $L$.  Since Nyquist electric field fluctuations are white noise in
  space, the $x$-dependence of $V$ behaves like a random walk in space
  $(\sim \sqrt{|x|})$, so that $Q (x) \sim |x|$.  For $\tauT \ll
  \TTh$, the potential seen during one traversal of the system is also
  white noise in time, i.e.\ $\delta_T (t) \to \delta (t)$.

  In the \emph{diffusive regime} ($ \tauT \ll t \ll \TTh $), 
    a random walk $x(t_1)$ of duration $t$ does not feel
    the boundaries, hence $|x(t_1)|\sim \sqrt{t_1}$. Thus ${\cal
    F}(t) \propto T \int_0^t \sqrt{t_1} dt_1 \propto T t^{3/2}$,
  reproducing the well-known result $\Tph \propto T^{-2/3}$
  \cite{AAK}.

In the \emph{ergodic regime} ($ \tauT \ll \TTh \ll  t$), the trajectory
fully explores the whole system, thus $|x(t_1)|\sim L$ instead, which
reproduces ${\cal F}(t) \propto T L t$ and $\Tph \propto T^{-1}$
\cite{SashaMontamb}.

We are primarily interested in the \textit{$0D$ regime}
reached at $T \ll \ETh $ ($ \TTh \ll \tauT \ll t$). In contrast to
the previous two regimes, a typical trajectory visits the vicinity of
any point $x$ in the interval $[0,L]$ several times during the time $
\tauT $ (see Fig.~\ref{Figure1}).  On time scales shorter than this
time $\tauT$ the potential is effectively frozen, so that the
broadened $\delta $-function in \Eq{Q-corr} saturates at its maximum,
$\delta_T(t)\to T$, and the variance of $V$ is of order $T^2 |x|$.
The phase picked up during $\tauT$ becomes $\varphi_\tauT = \tauT
\int_0^L \textrm{d}x \,p(x) V(x,t_1)$, where $p(x) \, \textrm{d}x$ is
the fraction of time the trajectory spends near $x$. Then only small
statistical deviations from the completely homogeneous limit, $p(x)=1/L$
(reached for $\tauT\rightarrow \infty$), yield a phase difference
between the two time-reversed trajectories. These deviations scale
like $\delta p\sim 1/\sqrt{\tauT}$, since the number of ``samples''
(i.e.\ of traversals of the system during time $\tauT$) effectively
grows with $\tauT$. Thus, setting $V\sim T \sqrt{|x|}$, we estimate
$\varphi_\tauT \sim \tauT L^{3/2} T/\sqrt{\tauT}$, so that
$\left<\delta \varphi_\tauT^2\right> \sim L^3 T^2 \tauT$. Adding up
the contributions from $t/\tauT$ independent time intervals
($t\gg\tauT$),  we find ${\cal F}(t)\propto L^3 T^2 t$, implying
$\Tph \sim T^{-2}$, characteristic of $0D$ systems \cite{SIA}. Thus,
when $\TTh$ becomes the smallest time scale, a \emph{dimensional
crossover} occurs and the system becomes effectively $0D$. 

The qualitative behavior of $\Tph$ in all three regimes also follows
upon extracting $\Tph$ selfconsistently from the standard
perturbative expression for the Cooperon selfenergy
\cite{Fukuyama1982,FlorianJan-1,FlorianJan-2}.  Inserting the usual
cutoffs $T$ and $1/\Tph$ for the frequency transfered between the
diffusive electrons and their Nyquist noise environment and
excluding the diffuson 0th mode via a cutoff at $1/L$ of the
transfered momentum, we have (omitting numerical prefactors):
\begin{equation}
\label{PertTheor}
\frac{1}{\Tph} \propto
    \frac{T}{g_1 L}
 \int_{1/\Tph}^{T} \!\!
    {\rm d}\omega
 \int_{1/L}^\infty \!\!
  \frac{D  \, {\rm d} q}{(Dq^2)^2 + \omega^2} , \quad  
 g_1 = \frac{h  \sigma_0}{e^2} \frac{A}{L} ,
\end{equation}
where $ g_1$ is the 1$D$ dimensionless conductance,
$ D = v_F l $ the 1$D$ diffusion constant, $ v_F $ the Fermi
velocity and $ \ell $ the mean free path.  Writing $\ETh = D/L^2$ this yields 
$\Tph \propto (g_1 / \sqrt{\ETh}T)^{2/3}$, $g_1/ T$ or $\ETh g_1/T^2$
for the diffusive ($ \tauT \ll \Tph \ll \TTh $), ergodic ($ \tauT \ll
\TTh \ll \Tph$) or $0D$ ($ \TTh \ll \tauT \ll \Tph$) regimes,
respectively, as above (with dimensionful parameters
reinstated). \Eq{PertTheor} illustrates succinctly that the modes
dominating dephasing lie near the infrared cutoff ($\omega \simeq
\Tph^{-1}$ or $\ETh$) for the diffusive or ergodic regimes, but near
the ultraviolet cutoff $\omega \simeq T$ for the $0D$ regime (which is why, in
the latter, the broadening of $\delta_T (t)$ becomes important).

\emph{Analytical results.}  We now turn to a  quantitative 
analysis \cite{details}. The diffuson in the ring geometry is
  \cite{MontBook}
\begin{equation}\label{RingDiffuson}
  Q(x) = \frac{ L_{\textrm{dw}} }{2} 
               \frac{\cosh\bigl([L-2|x|]/2 L_{\textrm{dw}}\bigr)}{\sinh(L/2 L_{\textrm{dw}})} \, ,
\end{equation}
where $ L_{\textrm{dw}} = \sqrt{D \Tdw} $, and $ x $ is the cyclic
coordinate along the ring. Terms of order $ \TTh/ \Tdw $ (small for an
almost isolated ring) do not change the parametric dependence of $
{\cal F} $ on $ T, \, t, \mbox{ and } L $ so that we neglect them
below, setting $\Tdw = \infty$ in
Eq.~(\ref{DecFunc}). Inserting
Eqs.(\ref{Q-corr},\ref{delta-T},\ref{RingDiffuson}) into
Eq.~(\ref{DecFunc}), the decay function $ {\cal F}_n(t) $ for a given
winding number $n$ can be calculated as in \cite{SashaMontamb}, but
replacing $\delta (t)$ by $\delta_T (t)$:
\begin{eqnarray}\label{DecFuncFinal}
  {\cal F}_n (t) & = &
    - \frac{4 \pi Tt}{g_1 L} \int_{0}^{1} {\rm d}u \, z(u) \, 
      \langle Q \rangle_{\textrm{crw}}(u) \, , \\
  \nonumber \langle Q \rangle_{\textrm{crw}}(u) & = &
    \frac{L}{2} \sum_{k=1}^{\infty} \frac{\cos(2 \pi k n u)}
{(\pi k)^2} e^{- (2 \pi k)^2 \, \ETh t \, u(1-u)} \, , \\
  \nonumber z(u) & = &
    - 2 \pi T t \, (1 - u) \, w(\pi T t u) + \!\! 
\int_{-\pi T t u}^{\pi T t u} \!\!\! {\rm d}v \, w(v) \, .
\end{eqnarray}
Eq.~(\ref{DecFuncFinal}) yields the following results for the
diffusive \cite{AAK,DiffusiveNonZeroN}, ergodic \cite{SashaMontamb}, and
$0D$ regimes at $ |n| \ll \ETh t $:
\begin{equation}\nonumber\refstepcounter{equation}\label{LimCases}
  {\cal F}_n (t) \simeq
   \left\{
     \begin{array}{lclr}
       \displaystyle
       \frac{\delta_{n,0} \pi^{3/2}}{2 g_1} \sqrt{\ETh}T t^{3/2} , 
       & \!\!\! & \tauT \ll t \ll \TTh; &
        \;\; (\theequation \text{a}) \\
       \displaystyle
       \frac{\pi T t}{3 g_1}, & \!\!\! & \tauT \ll \TTh \ll t; &
        \;\; (\theequation \text{b}) \\
       \displaystyle
       \frac{\pi^2}{270 \, g_1} \frac{T^2 t}{\ETh}, & \!\!\! & 
       \TTh \ll \tauT \ll t. & 
        \;\; (\theequation \text{c})
     \end{array}
   \right.
\end{equation}
Subleading terms in the three limiting cases (\ref{LimCases}a-c) are
of order $ {\cal O}\left[ (\tauT/t)^{1/2}, (t / \TTh)^{1/2} \right] $,
$ {\cal O}\left[ (\tauT / \TTh)^{1/2}, \TTh / t \right] $, and $ {\cal
  O}\left[ (\TTh / \tauT)^2, (\tauT / t) \right] $, respectively.
Note that the crossover temperatures
where $\tau_\varphi^{\rm diff} \simeq \tau_\varphi^{\rm erg}$ 
or $\tau_\varphi^{\rm erg} \simeq \tau_\varphi^{0D}$, 
namely  $c_1 g_1 \ETh$ or $c_2 \ETh$, respectively,
involve large  prefactors, $c_1 = 27/4 \simeq 7$
and $c_2 = 90/\pi \simeq 30$,
which should aid experimental efforts to reach the 0D regime.

\begin{figure}[t]
\ifpdf
  \includegraphics[width=\columnwidth]{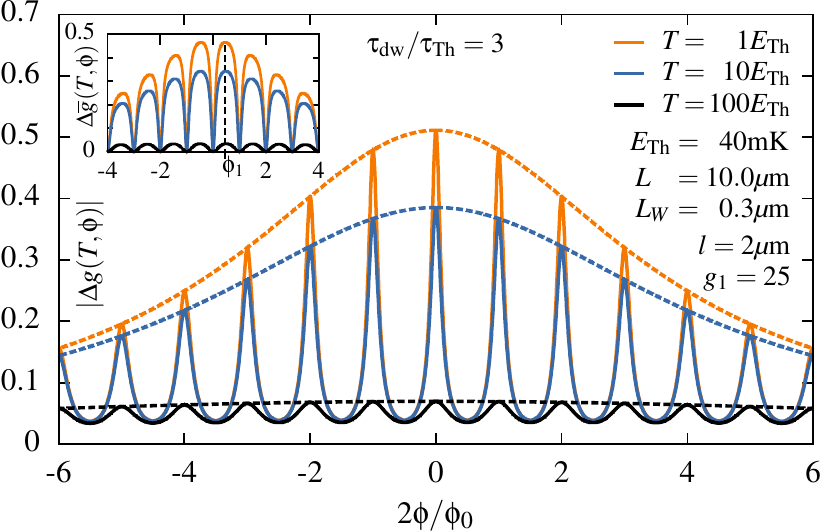}
\else
  \epsfig{file=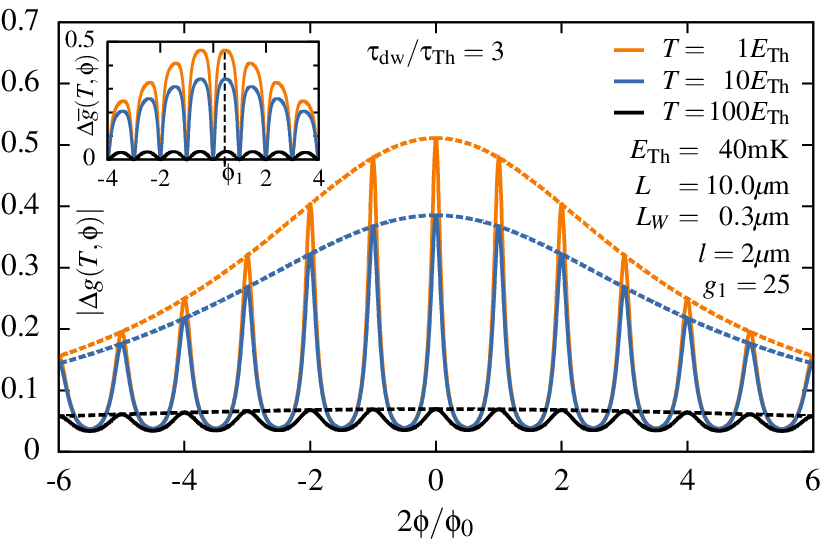,width=\columnwidth}
\fi
\caption{(Color online) The WL correction $ |\Delta g (T, \phi)|$
  (solid lines), its envelope $|\Delta g_\envelope (T, \phi)|$ (dashed
  lines), and their difference $\Delta \overline{g} = |\Delta
  g_\envelope| - |\Delta g| $ (inset), plotted as function of magnetic
  flux $ 2\phi/\phi_0 $, for three different temperatures.}
\label{WL-B}
\end{figure}

For a ring of rectangular cross section $A = L_W L_H$ and circumference
$L$, the Cooperon can then be written as
\begin{equation} \label{Coop-Sum}
  {\cal C}(t) \simeq
                 \sum_{n=-\infty}^{+\infty}
                 \frac{e^{-(nL)^2/4 D t}}{\sqrt{4 \pi D t}}
                 e^{ -t / \TphH - {\cal F}_n ( t ) - t / \Tdw }
              e^{i n \theta } \, , 
\end{equation}
with (restoring $\hbar$) $ \TphH = 9.5 (\hbar c / 
e H )^2  \times (l/D L_W^3) $ and $ \theta = 4 \pi \phi / \phi_0$,
where $\phi = \pi (L/2\pi)^2 H$ is the flux through the ring and 
$\phi_0 = hc/e$ \cite{MagnetoOsc,TauB,NetExper}.
Inserting Eqs.~(\ref{DecFuncFinal},\ref{Coop-Sum}) into
Eq.~(\ref{WL-sigma}) gives
the desired WL correction for the ring weakly coupled to leads.
The resulting value of $ \, | \Delta g(T, \phi) | \, $ increases with
decreasing $T$, Fig.~\ref{WL-B}, in a manner governed by $\Tph$: since only
trajectories with $|n| \lesssim 2\sqrt{t/\TTh}$ contribute, the
diffusive regime ($n$ restricted to 0) gives $|\Delta g|
\propto \sqrt{\Tph/ \TTh}/g_1 \propto (\ETh/g_1^2 T)^{1/3}$, whereas
the ergodic regime (sum on $n$ is $\propto t^{1/2}$) gives
$|\Delta g| \propto (\Tph / \TTh)/ g_1 \propto \ETh/T$, as long as $\Tph
< \TphH, \Tdw $.  With decreasing $T$, the growth of $| \Delta g (T,
\phi) | $ saturates towards $|\Delta g (0, \phi)|$ once $\Tph$ increases
past $\min (\TphH,\Tdw)$, with $|\Delta g (0, \phi)| -| \Delta g (T,
\phi)| \propto \Tph^{-1}$ vanishing as $T$ or $T^2$ in the ergodic
or $0D$ regimes, respectively.

\emph{Filtering out leads.}
For simplicity, above we did not model the leads explicitly.
In real experiments, however, $\Delta g$
is affected by dephasing in the leads, which might mask the
signatures of dephasing in  the confined region (the ring).  
This concern also applies to quantum dots connected to leads (cf.\
the $ \Tph \propto T^{-1} $-behavior observed in
\cite{DotExper}),
or finite-size effects in a network of disordered wires
\cite{NetExper}, where paths encircling a given unit cell might
spend significant time in neighboring unit cells as well (cf.\ $
T^{-1/3} $-behavior observed in \cite{NetExper,NetNote} at $ \Tph /
\TTh \geq 1 $).  The effect of leads can be filtered out
\cite{NetExper} by constructing from $|\Delta g (T, \phi)|$
its nonoscillatory envelope $| \Delta
g_\envelope (T, \phi) |$, obtained by
setting $\theta = 0$ in \Eq{Coop-Sum} while retaining $\TphH \neq
0$, and studying the difference
\begin{equation}
\label{Subtr}
    \Delta \overline{g}(T,\phi) 
= | \Delta g_\envelope (T,\phi)| - | \Delta g (T,\phi)| \; ;
\end{equation}
this procedure is illustrated in Fig.~\ref{WL-B}.  $\Delta
\overline{g}$ is dominated by paths with winding numbers $n \ge 1$
which belong to the ring. Contributions to $\Delta \overline{g}$
from Cooperons extending over both the ring and a lead will be
subleading for well-conducting leads with a small
contact-lead-contact return probability.  
Concretely, for $N$-channel point
contacts with conductance $\gcontact = N \transmission$,
this requires leads with dimensionless conductance $ \glead \gg N $
\cite{BrouwerPriv}.

\emph{Suggested Experiments.} To observe the predicted $1D$-to-$0D$
crossover experimentally, several conditions need to be satisfied.
Our theory
assumes (i) $ L \gg \ell \gg L_W \gg \lambda_{\text F} \, $ ($
\lambda_{\text F} $ is the Fermi wavelength). Ensuring that we stay in
the WL regime requires (ii) a large dimensionless conductance, $ g_1
\propto (\ell/L)(L_W L_H/\lambda_{\text F}^2) \gg 1$, and
(iii) a finite  $ \Tdw $ to limit the growth of $\Delta g$ with
decreasing $T$; choosing the limit, somewhat arbitrarily, as $ \Delta g 
\lesssim \frac{1}{2}$ at $T,H=0$ implies $\Tdw / \TTh \lesssim g_1 / 8$.
Estimating $\Tdw /  \TTh \simeq g_1 / \gcontact$,  this implies 
$8 \lesssim  \gcontact $, and thus the absence of Coulomb blockade. 
(iv) We also need $ \TTh \ll \Tdw $, or $\gcontact \ll g_1$, 
to ensure that trajectories with $ |n| \ge 1 $,
responsible for AAS oscillations, remain relevant. 
(v) To maximize the WL signal, the transmission per channel should
be maximal,  thus we suggest $\transmission \simeq 1$ and 
$N \simeq 10$.
(vi) The relevant temperature range, $[\Tmin, \Tmax ]$,
is limited from below by 
dilution refrigeration $(\Tmin \simeq 10{\rm mK})$
and from above by our neglect of phonons ($\Tmax \simeq 5$K).
(vii) The ring should be small enough that $c_2 \ETh \gtrsim \Tmin$.
 (viii) The interaction-induced dephasing rate
$\Tph^{-1}$, though decreasing with decreasing $T$, should 
for $T \simeq \Tmin$ not yet
be negligible compared to the $T$-independent rates $ \TphH^{-1} $
and $ \Tdw^{-1} $. These constraints can be met, e.g., with 
rings prepared from a $2D$  GaAs/AlGaAs heterostructure with 
$\lambda_{\text F} \approx 30$nm, $ v_F  \approx 2.5 \cdot 10^{5} $m/s, 
and $ g_1 = 4 \pi L_W l / \lambda_F L$,
by adjusting $ \, g_1 \, $ and
$ \ETh $ by  suitably choosing $L$ and $L_W$. 

\begin{figure}[t]
\ifpdf
  \includegraphics[width=\columnwidth]{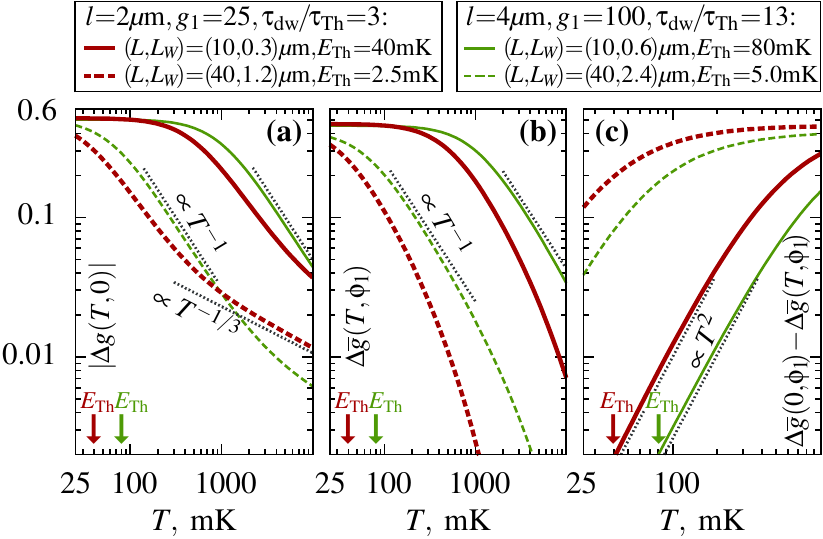}
\else
  \epsfig{file=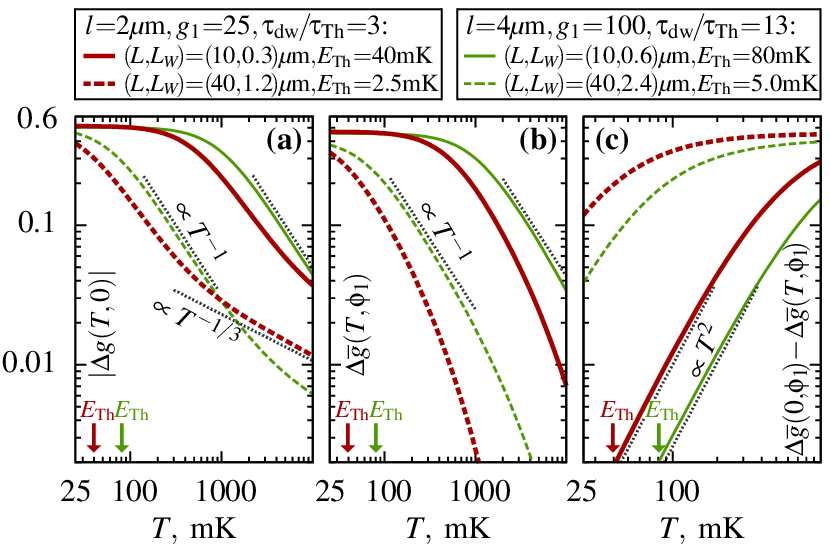,width=\columnwidth}
\fi
\caption{(Color online) $T$-dependence of (a) the WL correction at zero
  field, $ |\Delta g (T, 0)|$ and (b) at finite field
  with envelope subtracted, $\Delta \overline{g} (T, \phi_1)$; (c) the
  difference $\Delta \overline{g} (0, \phi_1) - \Delta \overline{g}
  (T, \phi_1)$, which reveals a crossover to $T^2$-behavior for $ T
  \ll 30 \ETh$. The flux $ \phi_1 $, which weakly depends on $T$,
  marks the first maximum of $ \Delta \overline{g}(T,\phi) $, see
  inset of Fig.~\ref{WL-B}.
         }
\label{WL-T}\vspace{-2mm}
\end{figure}

To illustrate this, numerical results for $|\Delta g|$ and $\Delta
\overline g$, obtained from Eq.(\ref{WL-sigma}) using experimentally
realizable parameters \cite{Baeuerle,Marcus:93,NetExper,2DEGexper},
are shown in Figs.~\ref{WL-B} and~\ref{WL-T} for several combinations
of $\ell$, $L$ and $L_W$.
The regime where $\Delta g$ exhibits
diffusive $T^{-1/3}$ behavior ($7 g_1 \ETh \ll T \ll \Tmax$) is visible
only for our smallest choices of both $g_1$ and $\ETh$
(Fig.~\ref{WL-T}a, heavy dashed line).
AAS oscillations in
$|\Delta g|$ and $\Delta \overline g$ (Fig.~\ref{WL-B}), which require
$\TTh \ll \Tph$, first emerge at the crossover from the diffusive to
the ergodic regime.  They increase in magnitude with decreasing $T$,
showing ergodic $T^{-1}$ behavior for $30 \ETh \ll T \ll 7
g_1 \ETh$ (Figs.~\ref{WL-T}a,b), and eventually saturate towards their
$T=0$ values, with $\Delta \overline{g} (0, \phi) - \Delta
\overline{g} (T, \phi)$ showing the predicted $0D$ behavior, $\propto
T^2$, for $T \lesssim 5 \ETh$, see Fig.~\ref{WL-T}c
(there $\Tph \gg \TTh$, i.e. dephasing is weak).

\emph{Conclusions.}  
The AAS oscillations of a quasi-$1D$ ring weakly coupled
to leads can be exploited to filter
out the effects of dephasing in the leads, thus offering
a way to finally observe, for $ T \lesssim  5 \ETh $, the elusive but
fundamental $0D$
behavior $\Tph \sim T^{-2}$.  This would allow
\emph{quantitative} experimental tests of the role of temperature as
ultraviolet frequency cutoff in the theory of dephasing.  

\acknowledgments We acknowledge helpful discussions with
B. Altshuler, C. B\"auerle, N. Birge, Y. Blanter, P. Brouwer,
L. Glazman, Y. Imry, V.  Kravtsov, J. Kupferschmidt, A. Mirlin,
Y. Nazarov, A. Rosch, D. Weiss and V. Yudson,
and support from the DFG through SFB TR-12, the Emmy-Noether program
and the Nanosystems Initiative Munich Cluster of Excellence;
from the NSF, Grant No. PHY05-51164;
and from the EPSRC, Grant No. T23725/01.


\begin{thebibliography}{30}
\bibitem{details} 
  Quantitative details, and an alternative 
  approach based on the kinetic equation,  will be presented elsewhere.
\bibitem{DiffusiveNonZeroN} Closed trajectories with $ n \ne 0 $
  contribute only weakly to $ \Delta \sigma $ at $ t \ll \TTh $, see
  Eq.(\ref{Coop-Sum}). That is why we do not include the case $ n \ne
  0 $ in Eq.(\ref{LimCases}a) \cite{SashaMontamb}.
\bibitem{NetNote}
  Note that $0D$ dephasing can not be observed in a network of
  well-connected unit cells, since trajectories are effectively not
  confined at $ \Tph / \TTh \gg 1 $.
\end{thebibliography}
\end{document}